\documentclass[pre,twocolumn,showpacs]{revtex4}
\usepackage{epsfig}
\usepackage{amssymb}
\begin{document}

\title{Water-wave gap solitons: An approximate theory 
and accurate numerical experiments} 
\author{V.~P. Ruban}
\email{ruban@itp.ac.ru}
\affiliation{Landau Institute for Theoretical Physics,
2 Kosygin Street, 119334 Moscow, Russia} 

\date{\today}

\begin{abstract}
It is demonstrated that a standard coupled-mode theory can successfully
describe weakly-nonlinear gravity water waves in Bragg resonance with a 
periodic one-dimensional topography.
Analytical solutions for gap solitons provided by this theory are in a
reasonable agreement with accurate numerical simulations of exact equations 
of motion for ideal planar potential free-surface flows, 
even for strongly nonlinear waves. 
In numerical experiments, self-localized groups of nearly standing 
water waves can exist up to hundreds of wave periods.
Generalizations of the model to the three-dimensional case are also derived.
\end{abstract}



\maketitle

\section{Introduction}

As we know from the nonlinear optics, specific self-localized waves can 
propagate in periodic nonlinear media, with a frequency inside a spectrum gap.
These waves are referred to as gap solitons (alternatively called 
Bragg solitons; see, e.g., Refs.\cite{CM1987,AW1989,CJ1989,ESdSKS1996,
PPLM1997,BPZ1998,RCT1998,CTA2000,IdS2000,CT2001,CMMNSW2008}).
Also in the field of Bose-Einstein condensation, gap solitons (GS) have been
known \cite{EC2003,PSK2004,MKTK2006}. Recently, it has been realized that
GS are also possible in water-wave systems \cite{R2008PRE-2}. In particular,
very accurate numerical experiments have shown that
finite-amplitude standing waves over a periodic one-dimensional topography are
subjected to a modulational instability which spontaneously produces Bragg
quasisolitons --- localized coherent structures existing for dozens of wave
periods. However, in the cited work \cite{R2008PRE-2}, no analytical approach
was presented. As a result, many important questions about water-wave GS 
were not answered, concerning their shape and stability. 
The present work is intended to clarify this issue, at least partly. 
More specifically, for a given periodic bottom profile with a spatial period 
$\Lambda$, we shall derive, in some approximation, 
coefficients for a standard model system of two coupled equations, 
describing evolution of the forward- and backward-propagating wave envelopes 
$A_\pm(x,t)$ 
(see, e. g., Refs.\cite{AW1989,CJ1989,RCT1998,CT2001}),
\begin{equation}\label{copuped_mode_equations}
i(\partial_t\pm V_g\partial_x)A_\pm=\Delta A_\mp+
(\Gamma_S|A_\pm|^2+\Gamma_X|A_\mp|^2)A_\pm,
\end{equation}
where $t$ is the time, $x$ is the horizontal coordinate in the flow plane,
and $A_\pm(x,t)$ are slow functions. Let at equilibrium the free surface 
be at $y=0$. Then elevation of the surface $y=\eta(x,t)$ is given by 
the following formula,
\begin{eqnarray}\label{elevation}
\eta(x,t)&=&\mbox{Re}\left[A_+e^{i\kappa x-i\omega_0(\kappa)t}
+A_-e^{-i\kappa x-i\omega_0(\kappa)t}\right]\nonumber\\
&&+\mbox{ higher-order terms in } \kappa A_\pm,
\end{eqnarray}
where $\kappa=2\pi/(2\Lambda)$ is the wave number corresponding to the main
Bragg resonance, $\omega_0(\kappa)=[g\kappa\tanh(h_0\kappa)]^{1/2}$ is the 
frequency at the gap center, $g$ is the gravity acceleration, and $h_0$ is an
effective depth of the water canal [definitely, $h_0$ is not a mean depth; 
more precisely it will be specified later by Eqs.(\ref{conformal}) and 
(\ref{bottom_parametrization})]. 
The coefficients in Eqs.(\ref{copuped_mode_equations}) are:
an effective group velocity $V_g=d\omega_0(\kappa)/d\kappa$, a half-width 
$\Delta$ of the frequency gap, a nonlinear self-interaction $\Gamma_S$,
and a nonlinear cross-interaction $\Gamma_X$. 

Generally, it is assumed in derivation of the above simplified standard model
that: (a) dissipative processes are negligible, 
(b) a periodic inhomogeneity is relatively weak (that is $\Delta\ll\omega_0$), 
(c) the waves are weakly nonlinear, (d) original (without inhomogeneity) 
equations of motion, when written in terms of normal complex variables 
$A_{\bf k}$, contain nonlinearities starting from the order three:
\begin{eqnarray}\label{original_equation}
i\dot A_{\bf k}&\approx&\omega_*({\bf k})A_{\bf k}
+\frac{1}{2}\int T({\bf k},{\bf k}_2;{\bf k}_3,{\bf k}_4)
A^*_{{\bf k}_2}A_{{\bf k}_3}A_{{\bf k}_4}\nonumber\\
&&\qquad\times \delta({\bf k}+{\bf k}_2-{\bf k}_3-{\bf k}_4)
\,d{\bf k}_2\,d{\bf k}_3\,d{\bf k}_4,
\end{eqnarray}
where $\omega_*({\bf k})$ is a linear dispersion relation in the absence of
periodic inhomogeneity (a weak inhomogeneity adds 
some small terms to the right hand side of Eq.(\ref{original_equation}); 
the most important effect arises from a term $\hat L A_{\bf k}$, where
$\hat L$ is a ``small'' linear non-diagonal operator). It is also required, 
(e) the coefficient $T({\bf k}_1,{\bf k}_2;{\bf k}_3,{\bf k}_4)$ 
of the four-wave nonlinear interaction should be a continuous function.
In application to water waves the requirements (d) and (e) mean that:
(i) all the second-order nonlinearities are assumed to be excluded by a 
suitable canonical transformation (the corresponding procedure is described, 
e.g., in Refs.\cite{K1994,Z1999});
(ii) the model system (\ref{copuped_mode_equations}) can be good only 
in the limit of relatively deep water, since on a finite depth the function 
$T({\bf k}_1,{\bf k}_2;{\bf k}_3,{\bf k}_4)$ is known to contain 
discontinuities which disappear on the infinite depth 
(see, e.g., Ref.\cite{Z1999}). Therefore we introduce a small parameter 
\begin{equation}
 \varepsilon\equiv\exp(-2\kappa h_0)\ll 1,
\end{equation}
and we consider in the main approximation only 
the principal effect of weak spatial periodicity, namely creation 
of a narrow frequency gap with $\Delta\sim\varepsilon\omega_0$
under the main Bragg resonance conditions. 
Then we imply a standard procedure
for obtaining approximate equations for slow wave envelopes, where  
the deep-water limit of 
$T({\bf k}_1,{\bf k}_2;{\bf k}_3,{\bf k}_4)$ is used for the coefficients 
$\Gamma_S\propto T(\kappa,\kappa;\kappa,\kappa)$ and
$\Gamma_X\propto 2T(\kappa,-\kappa;\kappa,-\kappa)$. Thus we neglect 
in the actual nonlinear wave interaction  some relatively small terms with
coefficients of order $\varepsilon$. 

Of course, the functions $A_\pm$ should be sufficiently ``narrow'' 
in the Fourier space, since dispersive terms proportional to second-order 
derivatives $\partial^2_x A_\pm$ are not included into the model.

After derivation of all the coefficients in section II, 
some known ``solitonic'' solutions of Eqs.(\ref{copuped_mode_equations}) 
will be compared to numerical results for exact hydrodynamic equations, 
with nearly the same initial conditions as in the solitons (in section III).
We shall see that very long-lived self-localized groups of standing 
water waves are possible. In some region of soliton parameters, 
water-wave GS exist up to hundreds of wave periods, until unaccounted by 
Egs.(\ref{copuped_mode_equations}) processes change them significantly.
In section IV we discuss some promising directions of further 
research, concerning three-dimensional generalizations of the coupled mode
equations. Some auxiliary calculations are placed in two Appendices.

\section{Coefficients of the model}

We start our consideration with a short discussion of conditions when 
dissipation due to bottom friction, caused by water (kinematic) viscosity 
$\nu$, is not important in wave dynamics. 
Obviously, a viscous sub-layer should be relatively thin in this case: 
$d_b\ll\Lambda$. In a nearly linear regime, 
a width of the sub-layer can be estimated as $d_b\sim(\nu/\omega)^{1/2}$, 
where $\omega\sim(g/\Lambda)^{1/2}$.
This gives us the following necessary condition for applicability of the
conservative theory: 
\begin{equation}
\Lambda^{3/4}g^{1/4}\nu^{-1/2}\gg 1.
\end{equation}
Generally speaking, one cannot exclude a possibility that in a strongly 
nonlinear regime the vorticity can sometimes be advected by a wave-produced 
alternating velocity field far away from the rigid bottom boundary.
Such vortex structures are typically generated near curved parts of the 
bed, and they can significantly interact with surface waves.
However, we assume this is not the case; otherwise, the problem becomes 
too complicated. Though we do not have simple criterion to evaluate 
influence of the bottom-produced vorticity, with 
$\Lambda\gtrsim 1$ m we still hope to be correct when neglecting 
water viscosity, as well as compressibility and surface tension. 
This allows us to exploit the model of purely potential free-surface 
ideal fluid flows, commonly used in the water wave theory. 

Since in this work we consider the case of relatively deep water, we can write
$\omega_0(\kappa)\approx \omega_*(\kappa)(1-\varepsilon)$, 
where $\omega_*(\kappa)=(g\kappa)^{1/2}$ is the frequency corresponding to the 
infinite depth. Later we will see that values $\varepsilon=0.01\dots 0.02$ 
are of the most interest.

\begin{figure}
\begin{center}
\epsfig{file=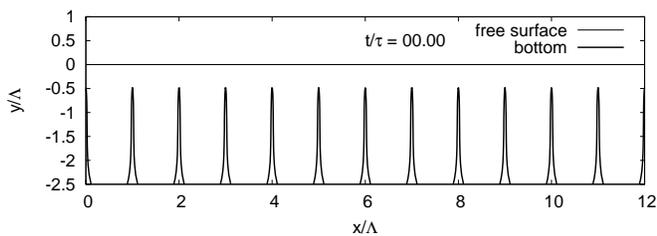,width=90mm}
\end{center}
\caption{Example I: unperturbed free surface and the bottom profile for
$2h_0\kappa=1.4\pi$, $D_0=0.95$, and $\varepsilon C=0.01229$
[see Eq.(\ref{B_explicit})].} 
\label{I_bottom} 
\end{figure}

Let us introduce conformal curvilinear coordinates $(\zeta_1,\zeta_2)$
determined by an analytic function ${\cal B}(\tilde \zeta)$, 
with $\tilde \zeta=\zeta_1+i(\zeta_2-h_0)$, so that
\begin{equation}\label{conformal}
x+iy={\cal B}(\tilde\zeta)=
\tilde\zeta-2\kappa^{-1}\sum_{n=1}^{\infty}
\beta_n\varepsilon^n\sin(2n\kappa \tilde \zeta),
\end{equation}
with real coefficients $\beta_n$. Without loss of generality, we assume 
$\beta_1 >0$. The unperturbed water surface $y=0$ corresponds 
to real values of $\tilde \zeta=\zeta_1-i0$, while at the bottom we have 
$\zeta_2=0$, and
\begin{equation}\label{bottom_parametrization}
X^{(b)}(\zeta_1)+iY^{(b)}(\zeta_1)={\cal B}(\zeta_1-ih_0)
\end{equation}
is a parametric representation of the bed profile, which can be highly
undulating (see, for example, Fig.\ref{I_bottom}).

In these conformal coordinates, a spectrum $\omega(\mu)$ of linear 
potential waves is determined through the following equation 
(compare to Ref.\cite{R2004PRE}, 
where an analogous approach but slightly  different notations were used):
\begin{equation}\label{linear_wave_spectrum_equation}
\left[\omega^2{\cal B}'(\zeta_1)-g\hat k\tanh(h_0\hat k)\right]
\Psi_\mu(\zeta_1)=0,
\end{equation}
with 
\begin{equation}
{\cal B}'(\zeta_1)=1-4\sum_{n=1}^{\infty}n\beta_n\varepsilon^n
\cos(2n\kappa \zeta_1).
\end{equation}
Here $[\hat k\tanh(h_0\hat k)]$ is a linear operator which is diagonal 
in Fourier representation: for any function 
$f(\zeta_1)=\int  f_k \exp(ik\zeta_1){dk}/{2\pi}$ we have
\begin{equation}
[\hat k\tanh(h_0\hat k)] f(\zeta_1)=
\int k\tanh(h_0 k)f_k e^{ik\zeta_1}{dk}/{2\pi}. 
\end{equation}
The eigenfunction $\Psi_\mu(\zeta_1)$ takes the following form,
\begin{equation}
\Psi_\mu(\zeta_1)=e^{i\mu\zeta_1}
\sum_{n=-\infty}^{+\infty} c_n e^{2in\kappa\zeta_1},
\end{equation}
with some coefficients $c_n$. With a given $\mu$, we have an infinite
homogeneous linear system of equations for $c_n$. Non-trivial solutions exist
for some discrete values $\omega_{(m)}(\mu)$. The first gap in the spectrum is
the difference between the two first eigenvalues at $\mu=\kappa$, that is
$2\Delta=\omega_{(2)}(\kappa)-\omega_{(1)}(\kappa)$. 
Approximately, for small $\varepsilon$ these eigenvalues are determined 
by the coefficient $\beta_1$ (compare to Ref.\cite{R2004PRE}),
\begin{equation}
\omega^2_{(1,2)}(\kappa)\approx
g\kappa\tanh(h_0\kappa)(1 \mp 2\varepsilon \beta_1).
\end{equation}
It should be noted that $\omega_{(1)}(\kappa)$ corresponds to
$\Psi^{(1)}_\kappa\approx\sin(\kappa \zeta_1)$, 
while $\omega_{(2)}(\kappa)$ corresponds to
$\Psi^{(2)}_\kappa\approx\cos(\kappa \zeta_1)$.
Thus, in the first order on $\varepsilon$, the half-width $\Delta$ of the
gap in the spectrum of linear waves is
\begin{equation}
\Delta\approx\omega_*(\kappa)\varepsilon\beta_1
\equiv \omega_*(\kappa)\tilde\Delta,
\end{equation}
where $\tilde\Delta=\varepsilon\beta_1\ll 1$ is a small dimensionless quantity.   
\begin{figure}
\begin{center}
\epsfig{file=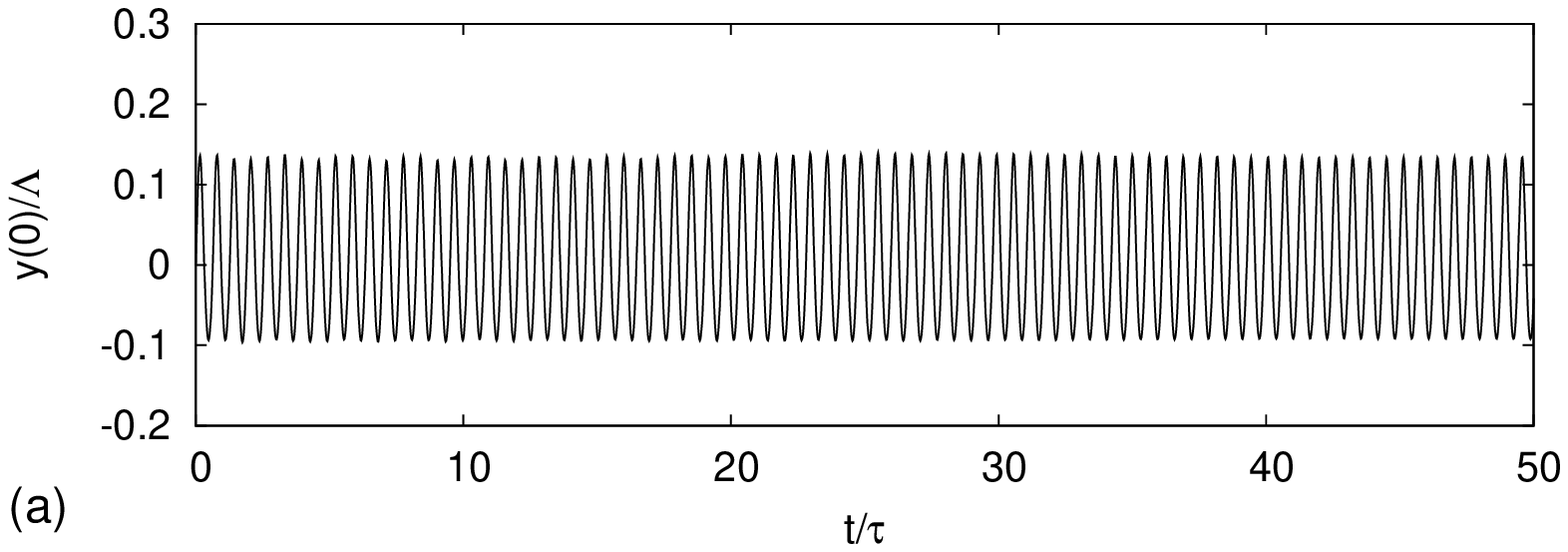,width=90mm}
\epsfig{file=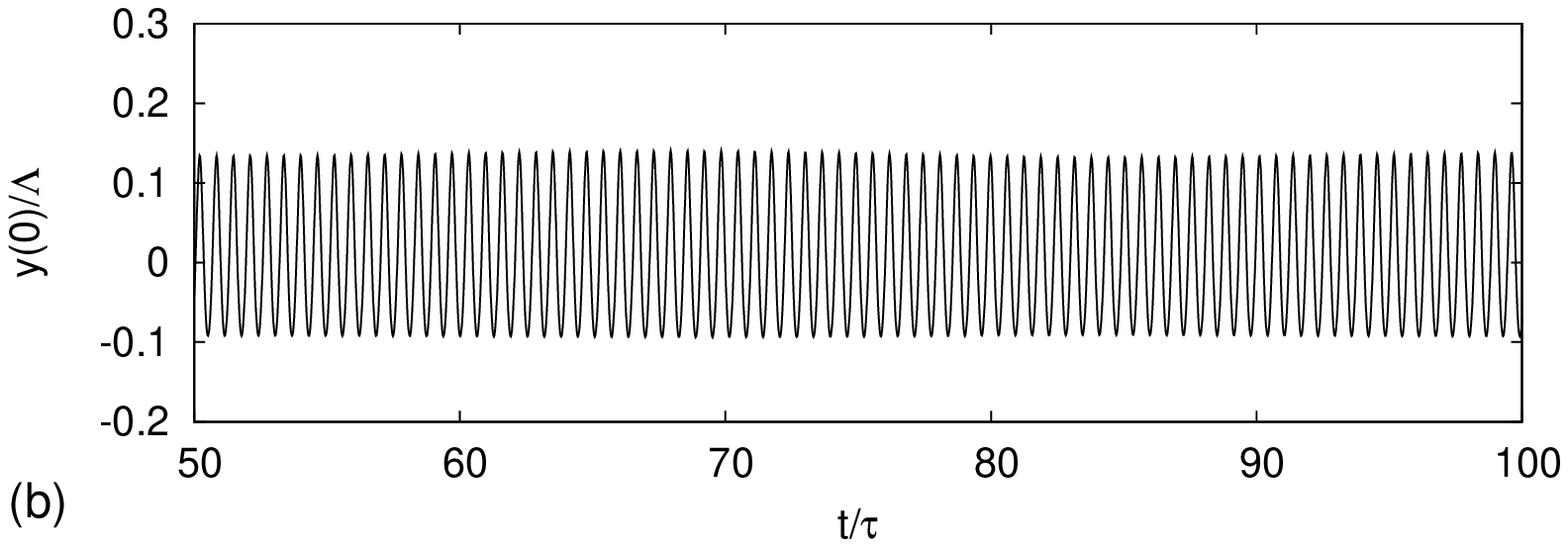,width=90mm}
\epsfig{file=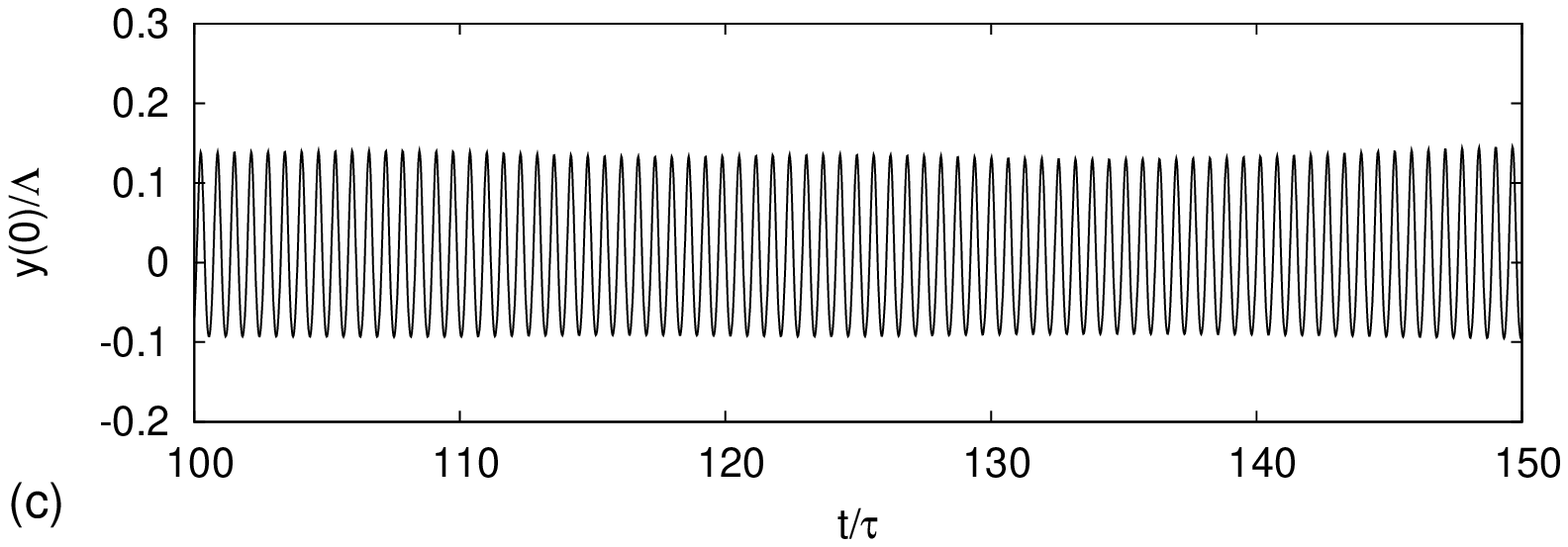,width=90mm}
\epsfig{file=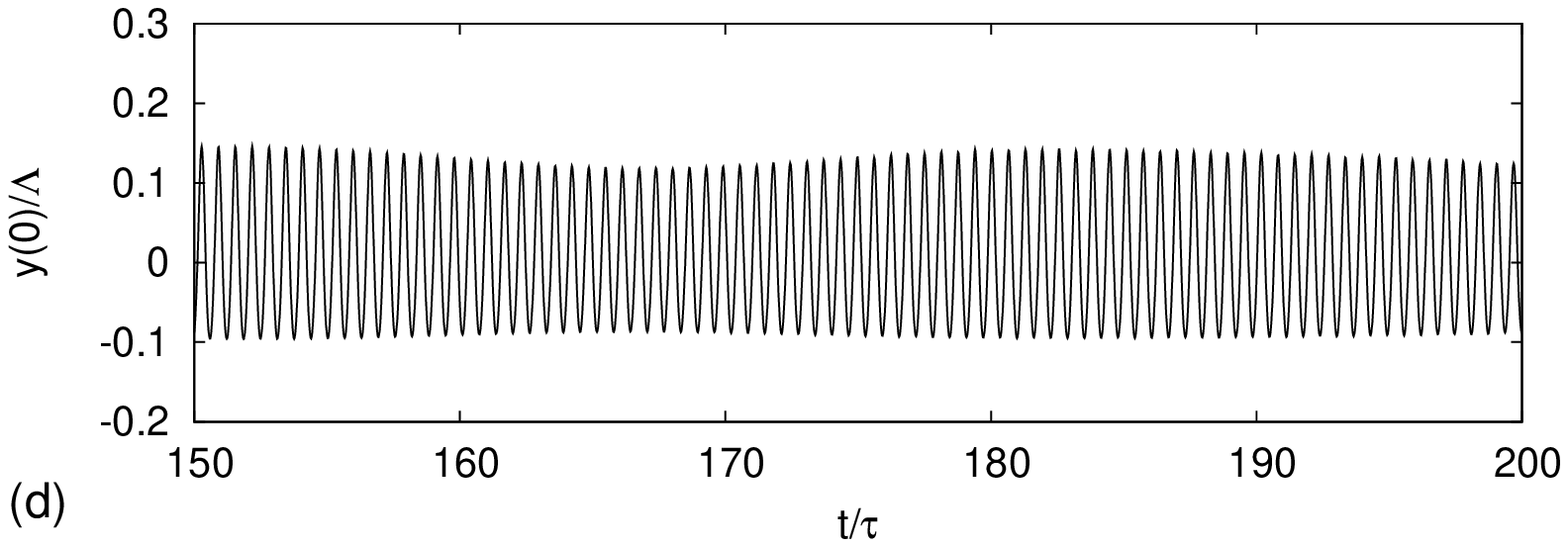,width=90mm}
\end{center}
\caption{Example I: free surface elevation at $x=0$ (at the center of GS) 
for $\delta=0.4$.} 
\label{I_Y_pi} 
\end{figure}
\begin{figure}
\begin{center}
\epsfig{file=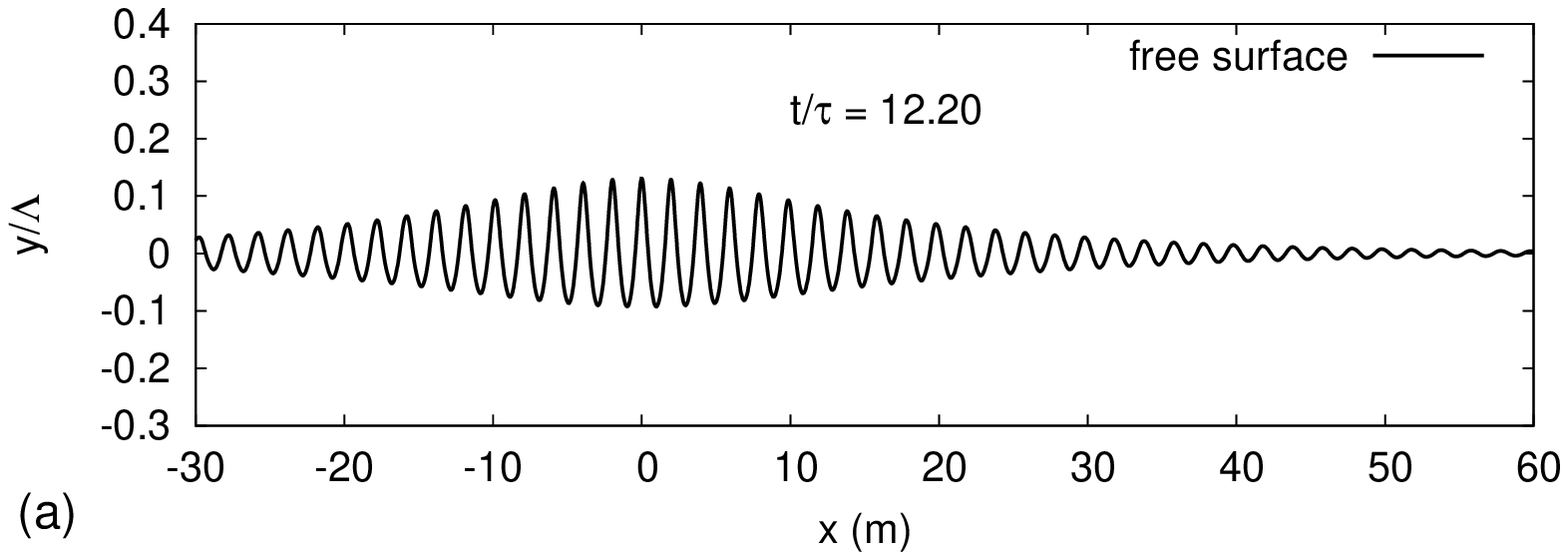,width=90mm}
\epsfig{file=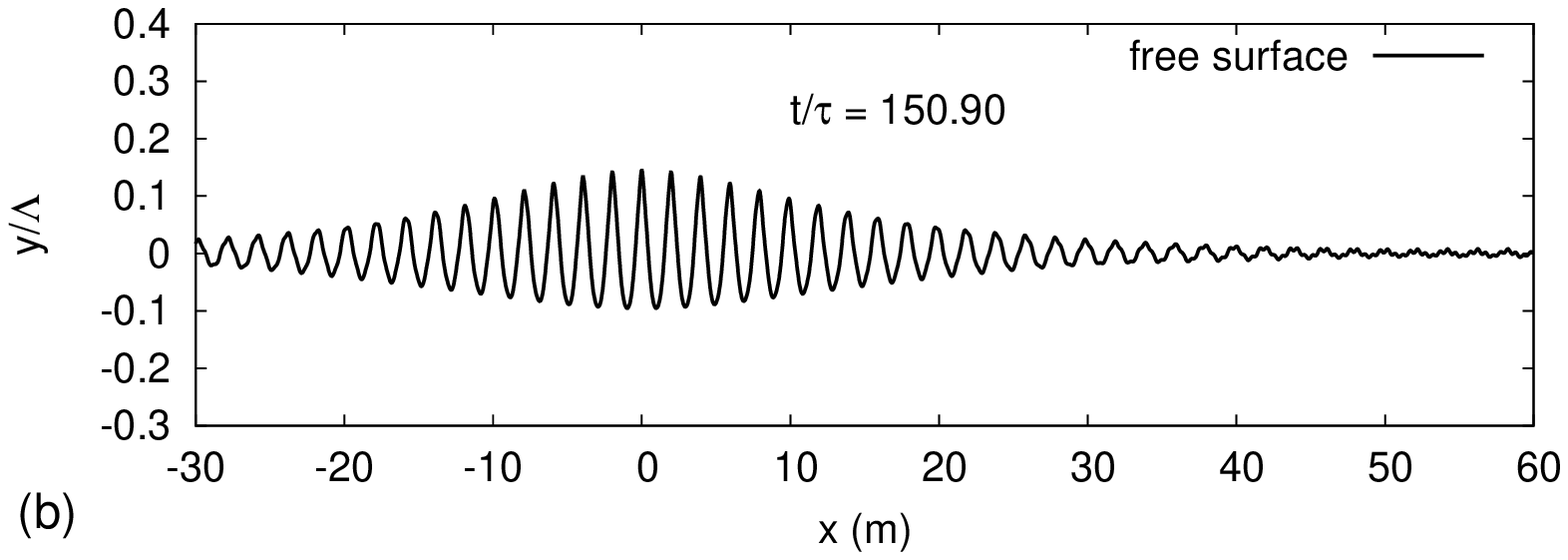,width=90mm}
\epsfig{file=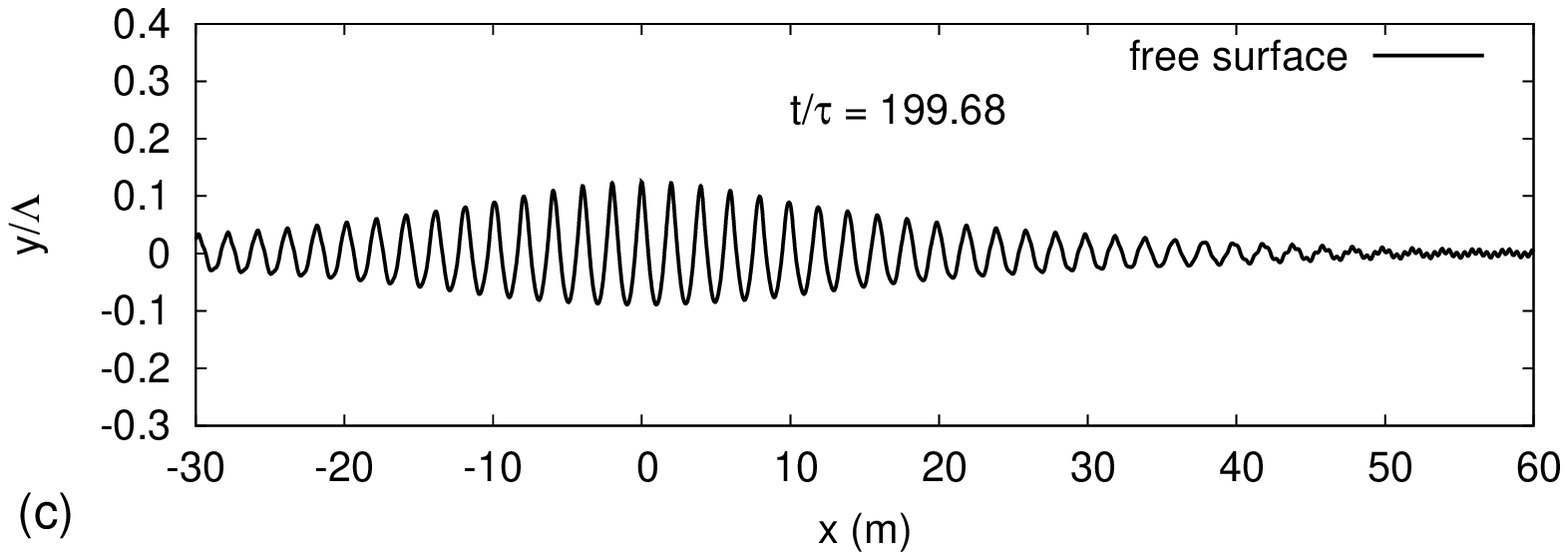,width=90mm}
\end{center}
\caption{Example I: free surface profiles for different time moments
when elevation at $x=0$ is at maximum.} 
\label{I_profiles} 
\end{figure}

As to the nonlinearity coefficients $\Gamma_S$ and $\Gamma_X$,
their values for the case of infinite depth (in other words, 
their zeroth-order approximations in $\varepsilon$) 
can be easily extracted from Ref.\cite{OOS2006}: 
\begin{equation}
\Gamma_S\approx\frac{1}{2}\omega_*(\kappa)\kappa^2,\qquad
\Gamma_X\approx-\omega_*(\kappa)\kappa^2.
\end{equation}
It should be noted, the cited work \cite{OOS2006} relies on results 
obtained earlier by Krasitskii \cite{K1994},
who calculated kernels of so-called reduced integrodifferential equation
for weakly nonlinear surface water waves 
(see also the paper by Zakharov \cite{Z1999}, and references therein).
It is important in many aspects that for deep-water waves the coefficients
$\Gamma_S$ and $\Gamma_X$ have the opposite signs, and their ratio is 
$\Gamma_S/\Gamma_X\approx -1/2$.

With the same zeroth-order accuracy, the group velocity is
\begin{equation}
V_g\approx \frac{1}{2}\frac{\omega_*(\kappa)}{\kappa}.
\end{equation}

Now all the coefficients have been derived, and the simplified 
coupled-mode equations for relatively deep water waves in Bragg resonance 
with a periodic bottom take the following explicit form:
\begin{eqnarray}
i\left(\frac{\partial_t}{\omega_*}+\frac{\partial_x}{2\kappa}\right)a_+=
\tilde\Delta\, a_- +\frac{1}{2}\left(|a_+|^2-2|a_-|^2\right)a_+,&&
\label{a_plus_eq}
\\
i\left(\frac{\partial_t}{\omega_*}-\frac{\partial_x}{2\kappa}\right)a_-=
\tilde\Delta\, a_+ +\frac{1}{2}\left(|a_-|^2-2|a_+|^2\right)a_-,&&
\label{a_minus_eq}
\end{eqnarray}
where $a_\pm(x,t)=\kappa A_\pm(x,t)$ are dimensionless wave amplitudes.
Analytical solutions are known for the above system (see 
\cite{AW1989,CJ1989,RCT1998,CT2001}), describing moving localized structures, 
the gap solitons. In the simplest case the velocity of GS is zero, 
and the solutions essentially depend on a parameter $\delta$, 
a relative frequency inside the gap ($-1<\delta<1$):
\begin{eqnarray}
a_\pm&=&\sqrt{I(x)}
\exp[{-i\delta\tilde\Delta \omega_*t +i\gamma_0 \pm i\varphi(x)}],\\
I(x)&=&\frac{4\tilde\Delta(1-\delta^2)}
{\cosh[4\tilde\Delta\sqrt{1-\delta^2}\,\kappa x] +\delta},\\
\varphi(x)\!&=&\!\arctan\!\left[\sqrt{\frac{1-\delta}{1+\delta}}
\tanh\left(2\tilde\Delta\sqrt{1-\delta^2}\,\kappa x\right)\right]\!.
\end{eqnarray}
These expressions correspond to purely standing, spatially localized
waves with frequency $\omega=(1-\varepsilon+\delta\tilde\Delta)\omega_*$
(concerning their stability, see Ref.\cite{RCT1998}, where, however, 
stability domains were presented for a different ratio $\Gamma_S/\Gamma_X$; 
there are some numerical indications that the above GS are stable
in a parametric interval $\delta_* < \delta < 1$, where a critical value 
$\delta_*\approx -0.4$).

It should be noted, one can hardly expect a detailed correspondence between
the very simple model (\ref{a_plus_eq}-\ref{a_minus_eq}) and the fully 
nonlinear dynamics, but just a general accordance sometimes is possible. 
In particular, the model does not describe nonlinear processes resulting 
in generation of short waves which take the wave energy away from a soliton, 
thus influencing its dynamics. The model is also not generally good to study
collisions between solitons, since wave amplitude can significantly increase 
in intermediate states.

\section{Numerical experiments}

\begin{figure}
\begin{center}
\epsfig{file=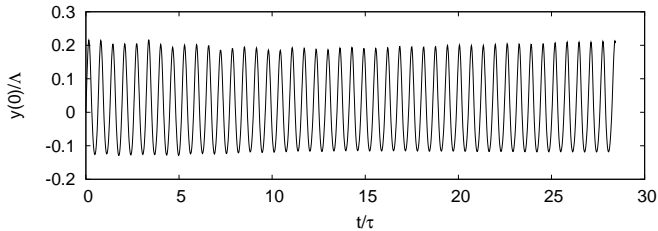,width=90mm}
\end{center}
\caption{Example II: free surface elevation at $x=0$  
for $2h_0\kappa=1.2\pi$, $D_0=0.7$, $\varepsilon C=0.022$, and $\delta=0.0$.} 
\label{II_Y_pi} 
\end{figure}
\begin{figure}
\begin{center}
\epsfig{file=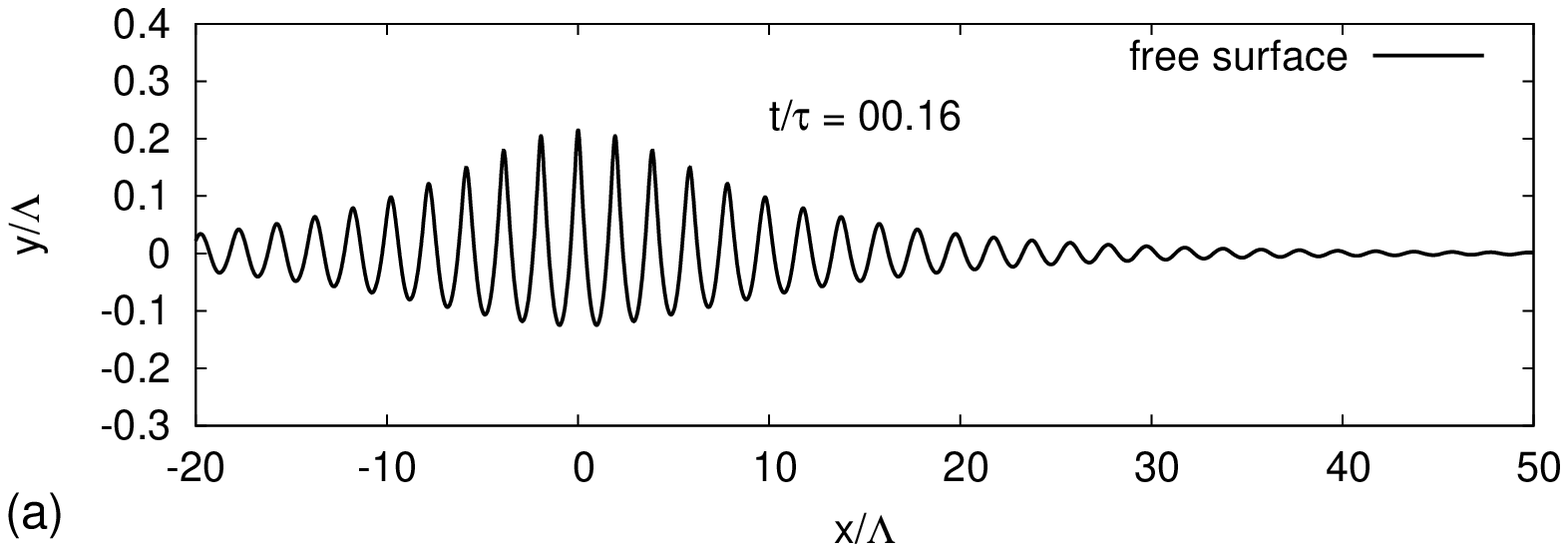,width=90mm}
\epsfig{file=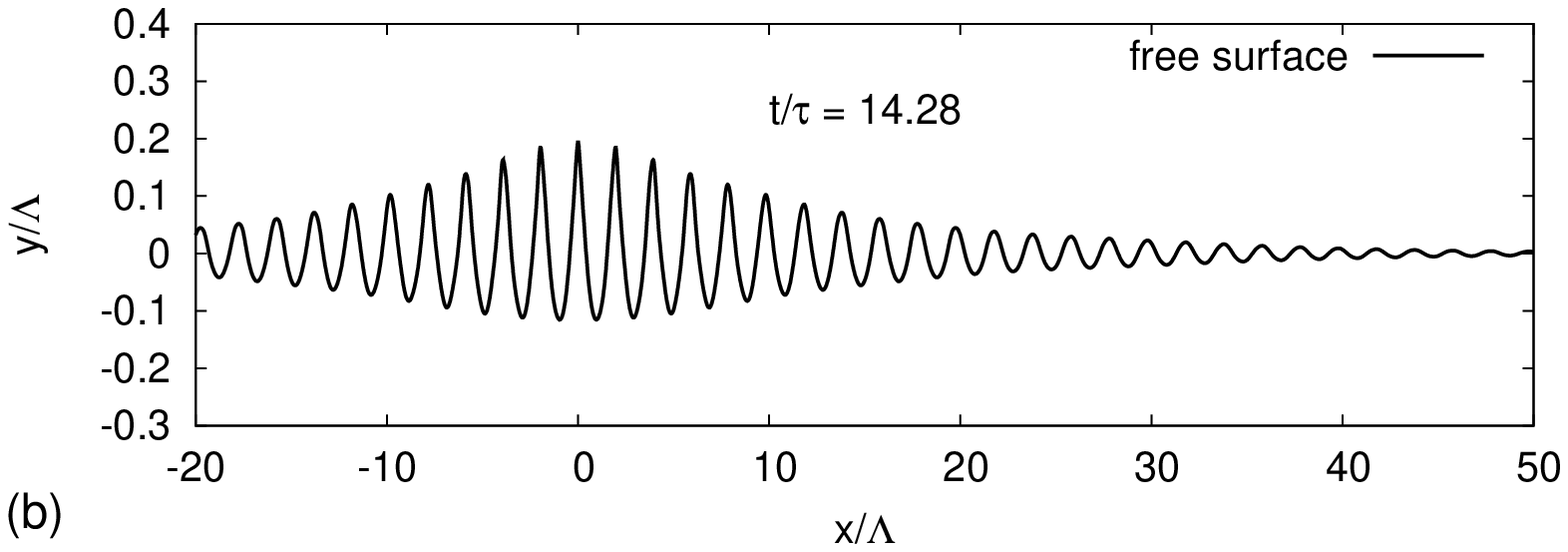,width=90mm}
\epsfig{file=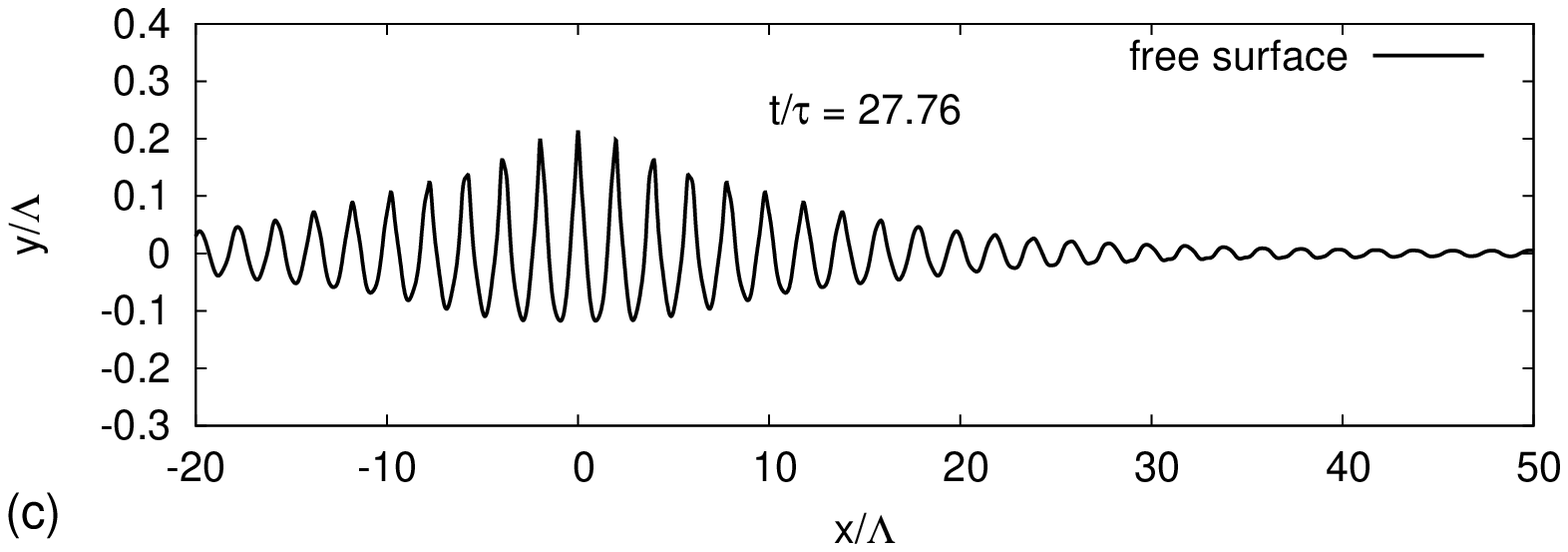,width=90mm}
\end{center}
\caption{Example II: free surface profiles for different time moments
corresponding to maximum elevation at $x=0$.} 
\label{II_profiles} 
\end{figure}
\begin{figure}
\begin{center}
\epsfig{file=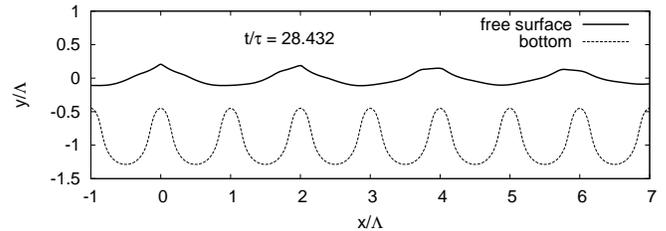,width=90mm}
\end{center}
\caption{Example II: formation of sharp wave crests over barriers.} 
\label{II_sharp_crests} 
\end{figure}
In order co compare the above approximate analytical solutions to 
nearly exact numerical solutions, we chose the following function 
${\cal B}(\tilde\zeta)$:
\begin{equation}\label{B_explicit}
{\cal B}(\tilde\zeta)=\tilde\zeta +\frac{iD_0}{\kappa}\ln\left(
\frac{1+\varepsilon C e^{2i\kappa \tilde\zeta}}
{1+\varepsilon C e^{-2i\kappa \tilde\zeta}}
\right),
\end{equation}
with real parameters $0< D_0<1$, and $0< C<1$. 
Hence, $\tilde\Delta=\varepsilon D_0 C<\varepsilon$. 
For $D_0$ and $C$ both close to 1, Eq.(\ref{B_explicit}) gives periodically 
arranged barriers (see, for example, Fig.\ref{I_bottom}). 
The barriers are relatively thin as $(1-D_0)\ll 1$, 
and relatively high as $C\to 1$. 
However, in numerical experiments with high-amplitude waves, 
a strong tendency  was noticed towards formation of sharp wave crests over 
very thin barriers (say, when $D_0=0.99$), already after a few wave periods.
With sharp crests, the conservative potential-flow-based model fails
(it is also clear that tops of narrow barriers must generate strong vortex
structures). 
Therefore we took $D_0=(0.7\dots 0.9)$ in most of our computations in order to 
have a smooth surface for a longer time.
Exact equations for ideal potential free-surface planar flows were simulated
(their derivation can be found in Ref.\cite{R2004PRE}, 
some generalizations are made in Refs.\cite{R2005PLA,R2008PRE}). 
As in Ref.\cite{R2008PRE-2}, we dealt
with dimensionless variables corresponding to $\tilde g=1$, $\tilde\kappa=100$.
The dimensionless time $\tilde t$ is then related to the physical time 
$t=\tau\tilde t$ by a factor $\tau=(100\Lambda/\pi g)^{1/2}$. 
For instance, the period of linear deep-water waves with the length 
$\lambda =2\Lambda$ is $T_*=(2\pi/\sqrt{100})\tau\approx 0.628\tau$.
At $t=0$, we set the horizontal free surface, while the
initial distribution of the surface-value velocity potential was
\begin{equation}
\psi_0(\zeta_1)= 2\kappa^{-3/2}\sqrt{I(\zeta_1)}
\cos[\kappa \zeta_1+\varphi(\zeta_1)]\approx\Psi_{GS}(\zeta_1),
\end{equation}
in accordance  with approximate relation $\eta_t\approx \kappa\psi$.

\begin{figure}
\begin{center}
\epsfig{file=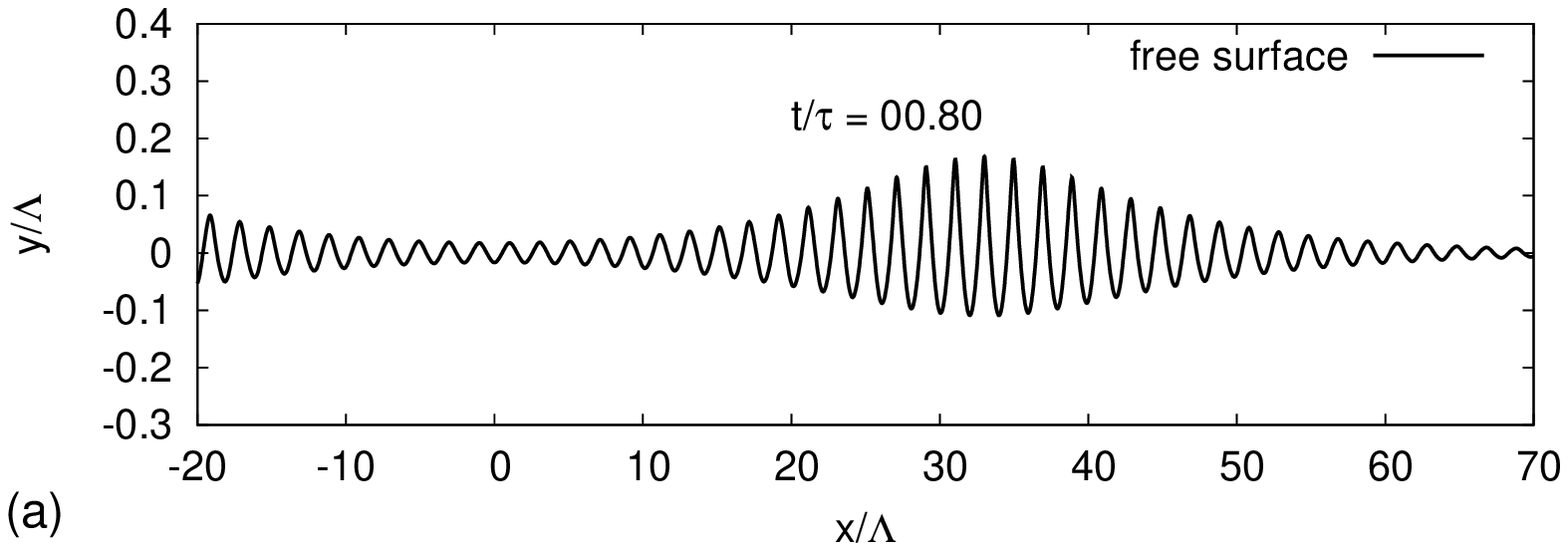,width=90mm}
\epsfig{file=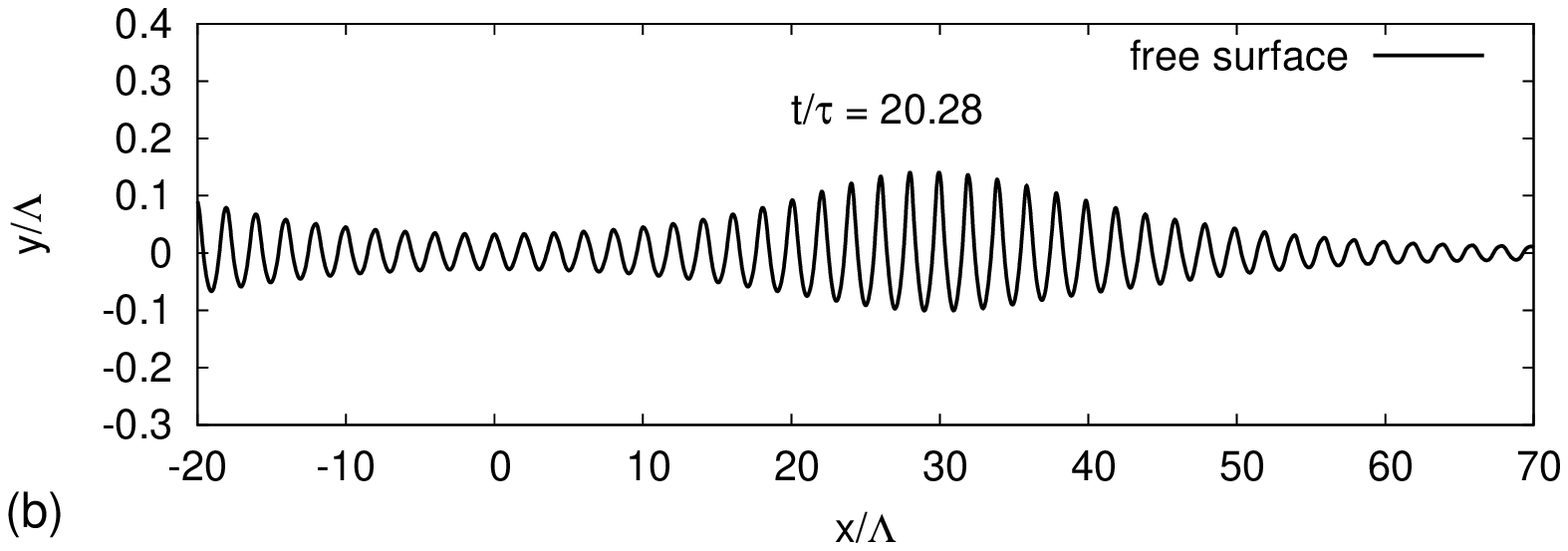,width=90mm}
\epsfig{file=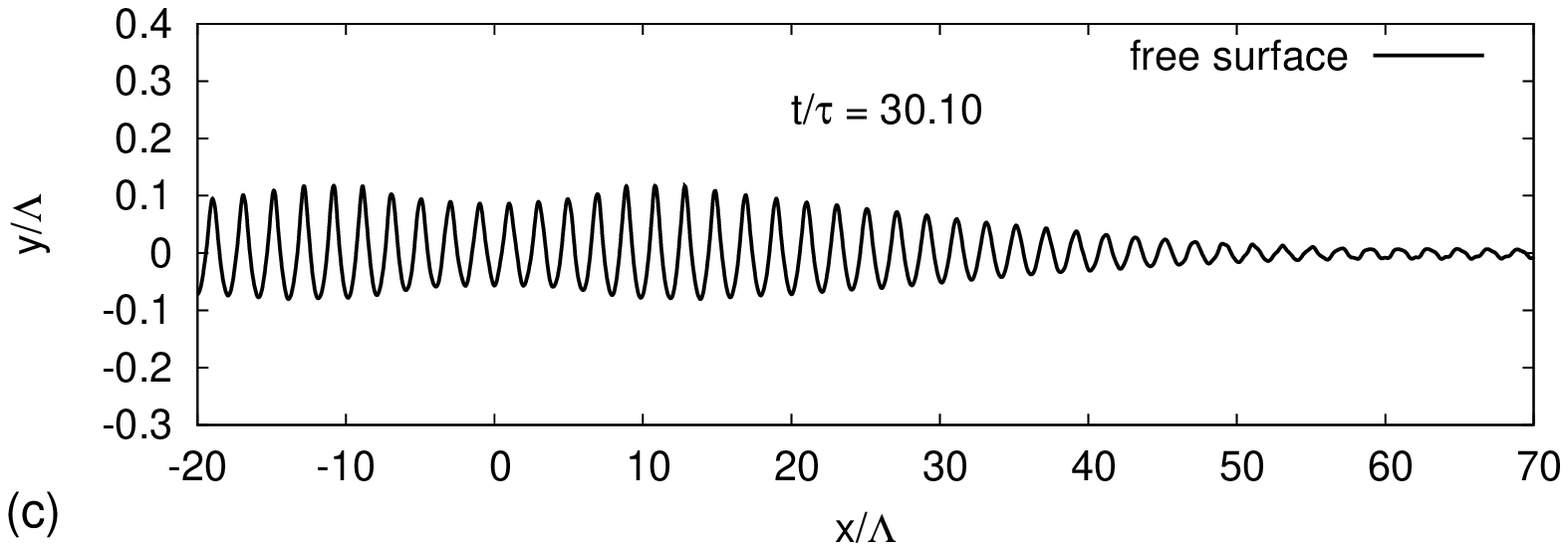,width=90mm}
\epsfig{file=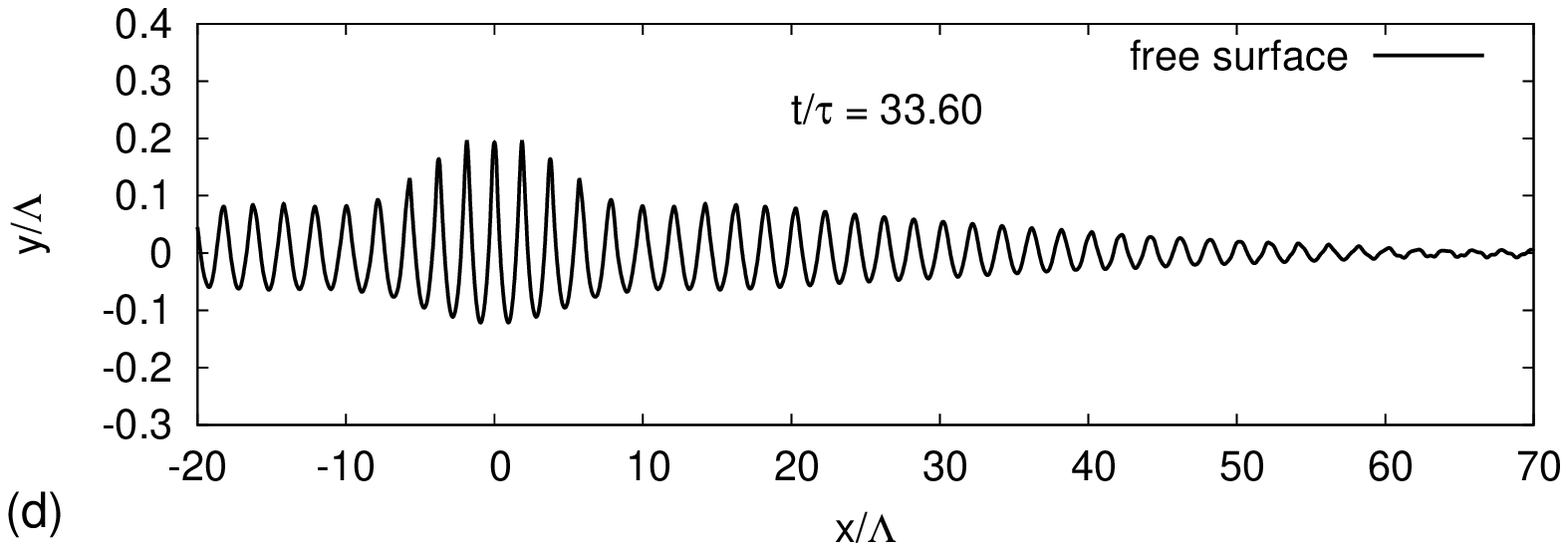,width=90mm}
\epsfig{file=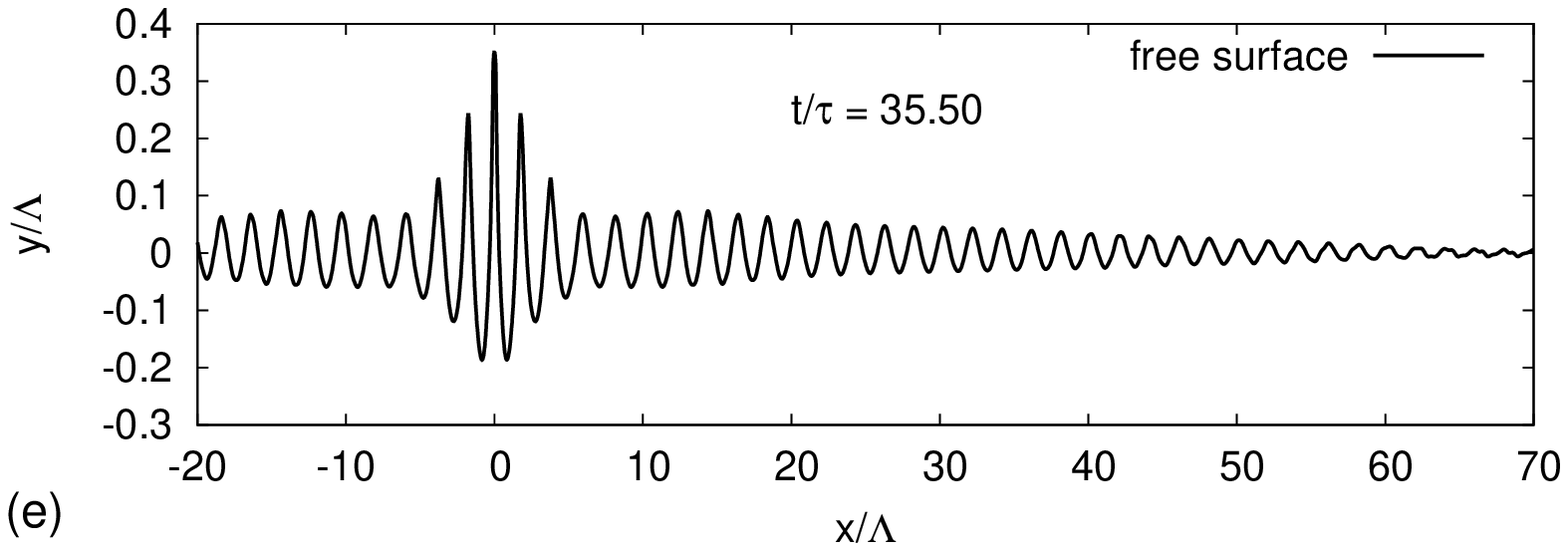,width=90mm}
\end{center}
\caption{Example III: interaction of two water-wave GS.} 
\label{III} 
\end{figure}
\begin{figure}
\begin{center}
\epsfig{file=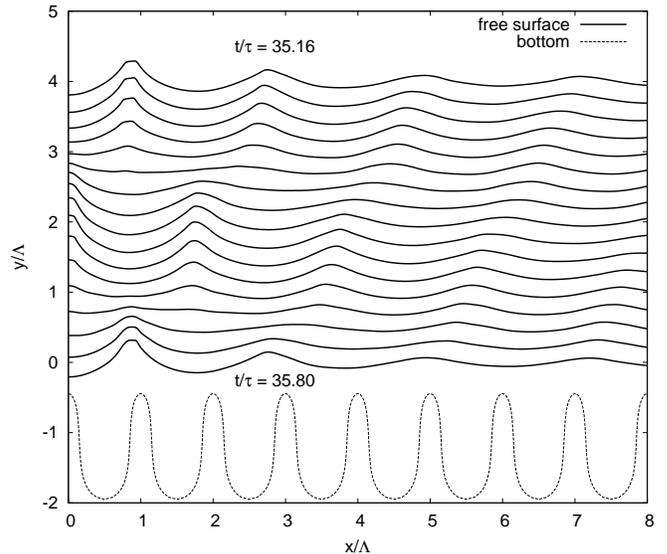,width=90mm}
\end{center}
\caption{Example III: highly nonlinear wave near $x=0$.
Here the free surface profiles are presented from 
$t/\tau=35.16$ to $t/\tau=35.80$ with the time interval $\Delta t/\tau=0.04$. 
The wave profiles, except the last one, are given vertically shifted for 
convenience. Also the bed shape is shown.} 
\label{III_final_stage} 
\end{figure}
Many simulations with different parameters were performed, and 
a very good general agreement was found between numerical and analytical 
results in the weakly-nonlinear case, that is for small steepness 
$s\equiv 2[I(0)]^{1/2}=4[\tilde\Delta(1-\delta)]^{1/2}\lessapprox 0.35$. 
So, with $2h_0\kappa=1.4\pi$, $D_0=0.95$, $\varepsilon C=0.01229$, 
and $\delta =0.4$ (example I), some noticeable deviations from the 
purely-standing-wave regime were observed only after
$t/\tau \gtrsim 120$ (see Figs.\ref{I_Y_pi}-\ref{I_profiles}).
In a real-world experiment it could be several minutes with $\Lambda\sim 1$ m.

What is interesting, even for larger $s$, up to $s\approx 0.48$, 
GS can exist for dozens of wave periods. 
A numerical example for such a relatively high-amplitude 
water-wave GS is presented in Figs.\ref{II_Y_pi}-\ref{II_profiles}, 
where $2h_0\kappa=1.2\pi$, $D_0=0.7$, $\varepsilon C=0.022$, and $\delta =0.0$
(example II).
In this simulation, there were 45 oscillations before sharp crests formation
(see Figs.\ref{II_Y_pi} and \ref{II_sharp_crests}). 
As to a further evolution of such GS, only in a real-world experiment 
it will be possible to get reliable knowledge about it, 
since various dissipative processes come into play.

Concerning water-wave GS with negative $\delta$, their behavior 
for $\delta > -0.4$ was found stable, while for $\delta \le -0.5$ 
the dynamics was unstable, and partial disintegration of GS was observed 
after a few tens of wave periods (not shown).
However, in some numerical experiments, the life time of GS was 
limited by the above mentioned process of sharp crest 
formation rather than by their own instability in frame of the model 
(\ref{a_plus_eq}-\ref{a_minus_eq}), at least with $s\gtrapprox 0.4$ 
(not shown). 

Finally, we would like to present an example of interaction of two  
GS (example III). The bed parameters are 
$2h_0\kappa=1.2\pi$, $D_0=0.7$, $\varepsilon C=0.023$. 
Both solitons initially had $\delta=0.4$ and they were separated 
by a distance $66\Lambda$. At $t=0$ we set the horizontal free surface and 
\begin{equation}
\psi_0(\zeta_1)\approx
\Psi_{GS}(\zeta_1-33\Lambda)+\Psi_{GS}(\zeta_1+33\Lambda).
\end{equation}
This numerical experiment also describes interaction of a single GS with a
vertical wall at $x=0$. Surface profiles for several time moments 
are shown in Figs.\ref{III}-\ref{III_final_stage}. We see that in this example 
the interaction between GS is attractive. 
They collide and produce a highly nonlinear and short wave group near $x=0$.

\section{3D generalizations and discussion}

In this work, coefficients of the standard model 
(\ref{copuped_mode_equations}) were derived for water-wave GS in
the approximation of relatively deep water. The frequency gap in this case is
small (of order $\varepsilon$) despite strong bed undulations. It seems that a
more general situation of intermediate depth cannot be described by this basic 
model, since an interaction of the main wave with a long-scale flow 
(``zeroth harmonics'') is then essential and should be included into equations.
At the formal level, this corresponds to the mentioned discontinuities of the
four-wave matrix element $T({\bf k}_1,{\bf k}_2;{\bf k}_3,{\bf k}_4)$
on a finite depth. Actually, in a finite-depth dynamics, 
three-wave interactions are more essential,
and therefore they cannot be removed efficiently by a 
weakly-nonlinear transformation. 
This is the main difference between the present third-order
theory and previously developed second-order theories (see, for example,
Ref.\cite{HM1987}).

So far we considered purely two-dimensional flows, with the
single horizontal coordinate $x$. Let us now introduce two important 
generalizations for three-dimensional flows. Below we only derive equations, 
but their detailed analysis will be a subject of future work.

In the first case, the bottom topography is still
one-dimensional, but we take into account weak variations of 
the wave field along the second  horizontal coordinate $q$, simply by adding 
dispersive terms, proportional to $\partial_q^2 a_\pm$, 
to the coupled-mode system, as written below:
\begin{equation}\label{2D_model_1Dbed}
\left(\frac{i\partial_t}{\omega_*}\pm
\frac{i\partial_x}{2\kappa}
+\frac{\partial_q^2}{4\kappa^2}\right)a_\pm
=\tilde\Delta\, a_\mp +\frac{1}{2}\left(|a_\pm|^2-2|a_\mp|^2\right)a_\pm.
\end{equation}
In this system, a near-band-edge approximation for the upper branch 
of the linear spectrum gives a 2D focusing nonlinear Schroedinger equation 
(NLSE). Thus, in a long-scale limit, the system (\ref{2D_model_1Dbed}) exhibits 
a tendency towards wave collapse which is known as a typical feature of 2D 
NLSE dynamics. 

\begin{figure}
\begin{center}
\epsfig{file=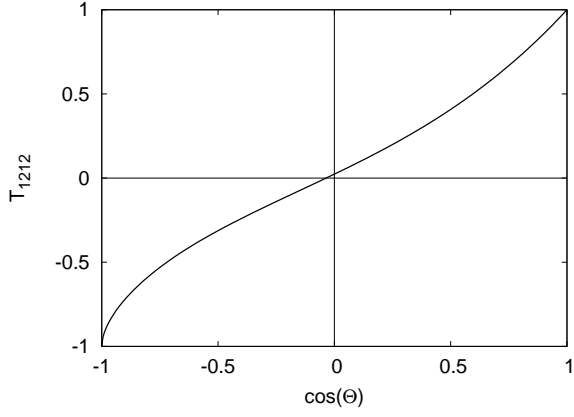,width=80mm}
\end{center}
\caption{A normalized matrix element 
$T({\bf k}_1,{\bf k}_2;{\bf k}_1,{\bf k}_2)$  for two wave vectors of equal 
length, with an angle $\Theta$ between them.} 
\label{T1212} 
\end{figure}

In the second case, a periodic bottom profile 
$Y^{(b)}(x,q)=-h+\chi(x,q)$ is essentially two-dimensional,
and in the horizontal Fourier-plane there are several pairs of Bragg-resonant 
wave vectors. For simplicity, we present below  equations for the case
when $\chi(x,q)$ has the symmetry of a square lattice, 
with equal periods $\Lambda$ in both horizontal directions $x$ and $q$:
\begin{equation}
\chi=\sum_{n_1 n_2} \alpha_{n_1 n_2}
[\cos(2n_1\kappa x+2n_2\kappa q)+\cos(2n_1\kappa x-2n_2\kappa q)],
\end{equation} 
where coefficients possess the symmetry $\alpha_{n_1 n_2}=\alpha_{n_2 n_1}$.
Let us consider interaction of two wave pairs having slow complex amplitudes 
$a_\pm(x,q,t)$ and $b_\pm(x,q,t)$, with the first pair corresponding to wave 
vectors $\pm{\bf p}_1=\pm(\pi/\Lambda)(1,1)$, and with the second pair 
corresponding to $\pm{\bf p}_2=\pm(\pi/\Lambda)(-1,1)$. It is important that
the absolute values are equal to each other:
$|{\bf p}_1|=|{\bf p}_2|=\sqrt{2}\kappa\equiv \varkappa$.
Again we will assume $\epsilon\equiv\exp(-2\varkappa h)\ll 1$.
It is convenient to use new horizontal coordinates:
\begin{equation}
x_1=\frac{q+x}{\sqrt{2}},\qquad x_2=\frac{q-x}{\sqrt{2}}.
\end{equation}
Elevation $y=\eta(x_1,x_2,t)$ of the free surface is then given by the formula
\begin{eqnarray}\label{2D_model_elevation}
\varkappa \eta&=&\mbox{Re}\Big\{e^{-i\Omega_0 t}\Big[
a_+e^{i\varkappa x_1}+a_-e^{-i\varkappa x_1}\nonumber\\
&&
+\,b_+e^{i\varkappa x_2}+ b_-e^{-i\varkappa x_2}\Big]\Big\} + \dots,
\end{eqnarray}
where 
$\Omega_0=[g\varkappa\tanh(h_0\varkappa)]^{1/2}\approx\Omega_*(1-\epsilon_0)$, 
with $\Omega_*=(g\varkappa)^{1/2}$ and $\epsilon_0=\exp(-2\varkappa h_0)$, 
and the dots correspond to higher-order terms (again, we should note that 
generally $h_0\not=h$). 
Approximate equations of motion for the amplitudes have the following form:
\begin{eqnarray}
i\left(\frac{\partial_t}{\Omega_*}\pm
\frac{\partial_{x_1}}{2\varkappa}\right) a_\pm=
\epsilon_1\, a_\mp +\epsilon_2\left[b_+ +b_-\right]
+\frac{\partial {\cal H}_{\rm{nl}}}{\partial a^*_\pm},&&
\label{a_eq}
\\
i\left(\frac{\partial_t}{\Omega_*}\pm
\frac{\partial_{x_2}}{2\varkappa}\right) b_\pm=
\epsilon_1\, b_\mp +\epsilon_2\left[a_+ +a_-\right]+
\frac{\partial {\cal H}_{\rm{nl}}}{\partial b^*_\pm},&&
\label{b_eq}
\end{eqnarray}
where small constants $\epsilon_1$ and $\epsilon_2$ depend on a given bed 
profile, $a^*_\pm$ and $b^*_\pm$ mean the complex conjugate quantities,
and the function ${\cal H}_{\rm{nl}}$ corresponds to nonlinear 
interactions.
Using an explicit expression from Ref.\cite{Z1999} for the deep-water
four-wave resonant interaction $T({\bf k}_1,{\bf k}_2;{\bf k}_3,{\bf k}_4)$, 
we have
\begin{eqnarray}
&&{\cal H}_{\rm{nl}}=\frac{1}{4}\left\{|a_+|^4+|a_-|^4
+|b_+|^4+|b_-|^4\right\}\nonumber
\\
&&-|a_+|^2|a_-|^2-|b_+|^2|b_-|^2\nonumber
\\
&&+\tau_\perp\left\{ |a_+|^2|b_+|^2+ |a_+|^2|b_-|^2
+|a_-|^2|b_+|^2+ |a_-|^2|b_-|^2\right\}\nonumber
\\
&&-\frac{3}{4}\left[a_+a_- b^*_+ b^*_- 
+ a^*_+ a^*_-  b_+ b_- \right],
\end{eqnarray}
where $\tau_\perp=T_{1212}(0)\approx 0.02346$ is a normalized value of 
the matrix element $T({\bf k}_1,{\bf k}_2;{\bf k}_1,{\bf k}_2)$  
for two  perpendicular wave vectors of equal length (see Fig.\ref{T1212}). 
Since  $\tau_\perp\ll 1$, we actually may neglect in ${\cal H}_{\rm{nl}}$ 
the terms proportional to $\tau_\perp$.

Unfortunately, it is hardly possible to find some analytical 
space-dependent solutions for the nonlinear system (\ref{a_eq})-(\ref{b_eq}), 
but it can be investigated by approximate methods.

The parameters  $\epsilon_0$, $\epsilon_1$, and $\epsilon_2$
can in principle be calculated from solution of a linearized problem
for water waves over a periodic 2D bed. An exact linearized equation 
for a surface value of the velocity potential
can be written in the following form: 
\begin{equation}\label{linear_3D}
(\omega^2/g-[\hat k\tanh(h\hat k)]-\hat N)\Psi_\omega({\bf r})=0,
\end{equation}
where ${\bf r}=(x,q)$ is a radius-vector in the horizontal plane, 
$\hat k=(\hat k_x^2+\hat k_q^2)^{1/2}$, 
while $\hat N$ is a self-conjugate  linear operator corresponding 
to a bottom inhomogeneity.
However, in three dimensions there is no compact form for $\hat N$,
valid with any bottom profile. 
At the moment, there are only approximate expressions 
$\hat N\approx \hat N_1+\hat N_2+\dots +\hat N_m$, obtained 
by expansion (up to a finite order $m$) of a vertical velocity at the 
level $y=0$ in powers of (relatively small) bottom deviation $\chi(x,q)$ 
from a constant level $y=-h$.
The linear self-conjugate  operators $\hat N_j$ have a general structure
\begin{equation}\label{N_S}
\hat N_j=[\cosh(h\hat k)]^{-1}\hat S_j [\cosh(h\hat k)]^{-1}, 
\end{equation}
with
\begin{equation}\label{S1}
\hat S_1=({\bf\nabla} \chi{\bf\nabla}),
\end{equation}
\begin{equation}\label{S2}
\hat S_2=-({\bf\nabla} \chi{\bf\nabla})
\left[\frac{\tanh(h\hat k)}{\hat k}\right]
({\bf\nabla} \chi{\bf\nabla}),
\end{equation}
\begin{eqnarray}
&&\hat S_3=({\bf\nabla}\chi{\bf\nabla})
\left[\frac{\tanh(h\hat k)}{\hat k}\right]
({\bf\nabla}\chi{\bf\nabla})\left[\frac{\tanh(h\hat k)}{\hat k}\right]
({\bf\nabla}\chi{\bf\nabla})
\nonumber\\
\label{S3}
&&\qquad\qquad+\left[\frac{1}{2}({\bf\nabla}\chi^2{\bf\nabla})
({\bf\nabla} \chi{\bf\nabla})
-\frac{1}{6}({\bf\nabla}\chi^3{\bf\nabla}){\bf\nabla}^2\right],
\end{eqnarray}
and so on, where ${\bf\nabla}$ is the horizontal gradient (see Appendix A). 

 Now we are going to calculate 
$\epsilon_0$, $\epsilon_1$, and $\epsilon_2$. Let us note that with 
$\epsilon\ll 1$ the four independent eigenfunctions in Bragg resonance are:
$\Psi_{\rm cc}\approx \cos(\kappa x)\cos(\kappa q)$, 
$\Psi_{\rm ss}\approx \sin(\kappa x)\sin(\kappa q)$,
$\Psi_{\rm cs}\approx \cos(\kappa x)\sin(\kappa q)$, and
$\Psi_{\rm sc}\approx \sin(\kappa x)\cos(\kappa q)$. Accordingly, we have for
the eigenfrequencies 
\begin{equation}
\omega^2_{\rm cc}/g\approx \varkappa\tanh(h  \varkappa)+
{\langle\Psi_{\rm cc}\hat N \Psi_{\rm cc}\rangle}/
{\langle \Psi^2_{\rm cc}\rangle},
\end{equation}
(where $\langle\dots\rangle$ mean the average value in the horizontal plane), 
and analogously for $\omega^2_{\rm ss}$ and 
$\omega^2_{\rm cs}=\omega^2_{\rm sc}$. Let us introduce short notations for
small quantities:
$
\nu_{\rm cc}\equiv{\langle\Psi_{\rm cc}\hat N \Psi_{\rm cc}\rangle}
/[{2\varkappa\langle \Psi^2_{\rm cc}\rangle}]\ll 1,
$
and similarly for $\nu_{\rm ss}$ and $\nu_{\rm cs}=\nu_{\rm sc}$. Then we have
approximate equalities,
\begin{eqnarray}
\omega_{\rm cc}&\approx&\Omega_*(1-\epsilon +\nu_{\rm cc}), \\
\omega_{\rm ss}&\approx&\Omega_*(1-\epsilon +\nu_{\rm ss}),\\
\omega_{\rm cs}=\omega_{\rm sc}&\approx&
\Omega_*(1-\epsilon +\nu_{\rm cs}).
\end{eqnarray}
These frequencies should be identified
with the eigenfrequencies of the linear part of system 
(\ref{a_eq})-(\ref{b_eq}), for space-independent solutions: 
\begin{eqnarray}
\omega_{(1,1,1,1)}\quad&=&\Omega_*[(1-\epsilon_0)+\epsilon_1 +2\epsilon_2]
=\omega_{\rm cc}, 
\\ 
\omega_{(1,1,-1,-1)}&=&\Omega_*[(1-\epsilon_0)+\epsilon_1 -2\epsilon_2]
=\omega_{\rm ss},  
\\
\omega_{(1,-1,1,-1)}&=&\Omega_*[(1-\epsilon_0)-\epsilon_1]=\omega_{\rm cs}, 
\\ 
\omega_{(1,-1,-1,1)}&=&\Omega_*[(1-\epsilon_0)-\epsilon_1]=\omega_{\rm sc}.
\end{eqnarray}
As the result, we obtain the required formulas for the model parameters:
\begin{eqnarray}
\epsilon_0&\approx&\epsilon
-\frac{1}{4}(\nu_{\rm cc}+\nu_{\rm ss}+2\nu_{\rm cs}),\\
\epsilon_1&\approx&\frac{1}{4}(\nu_{\rm cc}+\nu_{\rm ss}-2\nu_{\rm cs}),\\
\epsilon_2&\approx&\frac{1}{4}(\nu_{\rm cc}-\nu_{\rm ss}).
\end{eqnarray}
With Eqs.(\ref{N_S})-(\ref{S3}), calculation of 
$\nu_{\rm cc}$, $\nu_{\rm ss}$, and $\nu_{\rm cs}$ is straightforward 
if the function $\chi(x,q)$ contains a finite number of Fourier harmonics, 
for example 
\begin{equation}\label{bottom_example}
\chi=\frac{\alpha_1}{\kappa}(\cos 2\kappa x +\cos 2\kappa q)+
\frac{\alpha_2}{\kappa}\cos 2\kappa x \cos 2\kappa q.
\end{equation}
Moreover, since $\epsilon\ll 1$, it is possible to simplify the operators 
$\hat S_j$ by writing there $\hat k^{-1}$ instead of 
$[\tanh(h\hat k)/\hat k]$. By doing so and taking into account only $\hat N_1$
and $\hat N_2$, for the bottom profile (\ref{bottom_example}) we obtain 
approximately
\begin{eqnarray}
\epsilon_2&\approx&-\frac{2\epsilon}{\sqrt{5}}\alpha_1\alpha_2,\\
\epsilon_1&\approx&\frac{\epsilon}{\sqrt{2}}\alpha_2,\\
\epsilon_0&\approx&\epsilon\left[1+\frac{8}{\sqrt{5}}\alpha_1^2
+\left(1+\frac{1}{2\sqrt{5}}\right)\alpha_2^2\right].
\end{eqnarray}

Thus, the expansion $\hat N\approx \hat N_1+\hat N_2+\dots +\hat N_m$ 
is certainly useful for analysis of the case $|\nabla \chi|\ll 1$, 
but it can hardly be valid for a strongly undulating bed.
It should be noted that a global representation of the velocity potential 
in the form (\ref{velocity_potential}) (see Appendix A) 
is questionable in the general case.
Derivation of $\hat N$ for arbitrary $|\nabla b|$, assuming $\epsilon\ll 1$, 
is an interesting open problem. 

It is worth noting that an explicit (though approximate) 
form of operator $\hat N$ allows us 
to derive weakly nonlinear equations of motion for water waves 
over a nonuniform 2D bottom. For example, the Hamiltonian functional
(it is the kinetic energy ${\cal K}$ plus the potential energy 
$(g/2)\int\eta^2 d^2{\bf r}$) 
up to the 4th order in terms of the canonically conjugate variables 
$\eta({\bf r}, t)$ and $\psi({\bf r},t)$ is written below:
\begin{eqnarray}
{\cal H}&\approx& \frac{1}{2}\int\left\{
\psi\hat K\psi +g\eta^2+ \eta \left[(\nabla \psi)^2
-(\hat K\psi)^2\right]\right\} d^2{\bf r}\nonumber\\
&&+\frac{1}{2}\int\left[
\psi\hat K\eta \hat K\eta\hat K\psi 
+\eta^2(\hat K\psi)\nabla^2\psi \right] d^2{\bf r},
\label{H_up_to_4}
\end{eqnarray}
where $\hat K\equiv[\hat k\tanh(h\hat k)+\hat N]$ (see Appendix B). 
It is interesting to note  that the bottom inhomogeneity comes into 
the Hamiltonian through the definition of operator $\hat K$ only. 
For $\hat N =0$, it coincides with the previously known fourth-order 
Hamiltonian for water waves on a uniform depth 
(see, e.g., Ref.\cite{Z1999}, and references therein).
It is also clear that coupled-mode system (\ref{a_eq})-(\ref{b_eq})
corresponds to the case $\hat K\approx \hat k$, when 
the difference $[\hat K-\hat k]$ is neglected in the third- and fourth-order 
parts of the Hamiltonian, but it is kept in the second-order part.
The functional ${\cal H}\{\eta,\psi\}$ determines canonical 
equations of motion,
\begin{eqnarray}
\eta_t&=&\frac{\delta{\cal H}}{\delta\psi}\approx
\hat K\psi-(\nabla \eta\nabla)\psi-\hat K \eta \hat K\psi\nonumber\\
&+&\!\hat K\eta \hat K\eta\hat K\psi
+\frac{1}{2}\left[\hat K\eta^2\nabla^2\psi
+\nabla^2\eta^2\hat K\psi\right],
\label{y_t_2nd_order}\\
-\psi_t&=&\frac{\delta{\cal H}}{\delta \eta}\approx g\eta 
+\frac{1}{2} \left[(\nabla \psi)^2-(\hat K\psi)^2\right]\nonumber\\
&+&(\hat K\psi)\hat K(\eta\hat K\psi)+\eta(\hat K\psi)\nabla^2\psi.
\end{eqnarray}
Numerical simulation of these cubically nonlinear equations, 
with $\hat N\not=0$, will be an important subject of future research. 

Further analytical and computational work is also needed 
to investigate formation of vortex structures near the bottom boundary 
and to evaluate their influence on the free surface dynamics.
In any case, the present results, based on the 2D purely potential theory, 
are deserving much attention. Moreover,
the author hopes that in a future real-world experiment all the
mentioned dissipative processes will not be able to destroy water-wave GS for a
sufficiently long time. Instead,  with vortices and breaking wave crests, the
predicted phenomenon of standing self-localized water waves
over a periodic bed will be found even more rich, interesting, and beautiful.

{\it Acknowledgments.}
These investigations were supported 
by RFBR grant No. 06-01-00665,  by RFBR-CNRS grant No. 07-01-92165,
by the ``Leading Scientific Schools of Russia'' grant No. 4887.2008.2,
and by the Program ``Fundamental Problems of Nonlinear Dynamics'' 
from the RAS Presidium.

\begin{widetext}
\section*{Appendix A. Expansion of operator $\hat N$}

The expansion of $\hat N$ in powers of $\chi$ is easily obtained from the
integral representation of the velocity potential
\begin{equation}\label{velocity_potential}
\Phi({\bf r},y)=\int\left[\phi_{\bf k}\frac{\cosh k(y+h)}{\cosh kh}
+f_{\bf k}\frac{\sinh ky}{k}
\right]e^{i{\bf kr}}\frac{d^2{\bf k}}{(2\pi)^2},
\end{equation}
where $\phi_{\bf k}$ is the Fourier transform of the velocity potential at
$y=0$, and $f_{\bf k}$ is the Fourier transform of an unknown function 
$f({\bf r})$ which should be determined by substitution of 
Eq.(\ref{velocity_potential}) into the bottom boundary condition 
\begin{equation}\label{bottom_boundary_condition}
\left[{\partial\Phi}/{\partial y}-{\bf\nabla}\chi\cdot
{\bf\nabla}\Phi\right]\big|_{y=-h+\chi}=0.
\end{equation}
The resulting integral equation can be represented as follows, 
\begin{equation}
-{\bf\nabla}\cdot\int
f_{\bf k}\frac{i{\bf k}\cosh [k(\chi({\bf r})-h)]}{k^2}
\exp({i{\bf kr}})\frac{d^2{\bf k}}{(2\pi)^2}
\label{equation_for_f} 
-{\bf\nabla}\cdot\int\phi_{\bf k}
\frac{i{\bf k}\sinh [k\chi({\bf r})]}{k\cosh [kh]}
\exp({i{\bf kr}})\frac{d^2{\bf k}}{(2\pi)^2}=0.
\end{equation} 
It can be formally solved for $f({\bf r})$ by expanding
Eq.(\ref{equation_for_f}) in powers of $\chi$ and assuming $f=f_1+f_2+\dots$.
For instance, equation (\ref{equation_for_f}) with the third-order accuracy 
is written below:
\begin{equation}
\left\{1+({\bf\nabla}\chi{\bf\nabla}) 
\left[\frac{\tanh (h\hat k)}{\hat k} \right]
-\left({\bf\nabla}\frac{\chi^2}{2}{\bf\nabla}\right) \right\}
[\cosh(h\hat k)] f
=\left[({\bf\nabla}\chi{\bf\nabla})-
\left({\bf\nabla}\frac{\chi^3}{6}{\bf\nabla}\right){\bf\nabla}^2\right]
[\cosh(h\hat k)]^{-1}\phi.
\end{equation}
As the result, be obtain an approximate solution 
$f\approx(\hat N_1+\hat N_2+\hat N_3)\phi$, where
the operators $\hat N_1$, $\hat N_2$, and $\hat N_3$ are given by 
Eqs.(\ref{N_S})-(\ref{S3}).
A linearized system describing the free-surface dynamics is
\begin{equation}
-\psi_t=g\eta,\qquad \eta_t=[\hat k\tanh(h\hat k)]\psi +f,
\end{equation}
where $\psi({\bf r})=\Phi({\bf r},\eta({\bf r}))$ is a surface value of
the velocity potential (in the linear approximation $\psi\approx\phi$).
It gives us the equation (\ref{linear_3D}) for eigenfunctions $\Psi_\omega$
corresponding to some fixed frequency $\omega$.

\section*{Appendix B. Hamiltonian of water waves up to 5th order} 

An approximate Hamiltonian of water waves  can be easily derived by
writing the kinetic energy of potential three-dimensional 
motion of an ideal fluid in the following form,
\begin{eqnarray}
{\cal K}&=&\frac{1}{2}\int d^2{\bf r}\!\!\!
\int\limits_{-h+\chi({\bf r})}^{\eta({\bf r})}\!\!
\left[({\partial\Phi}/{\partial y})^2
+({\bf\nabla}\Phi)^2\right]dy
=\frac{1}{2}\int \psi
\left[{\partial\Phi}/{\partial y}-{\bf\nabla}\eta\cdot
{\bf\nabla}\Phi\right]\big|_{y=\eta}\,d^2{\bf r}\nonumber\\
&=&\frac{1}{2}\int \nabla\psi\cdot\int\left[
\phi_{\bf k}\frac{i{\bf k}\sinh [k(\eta({\bf r})+h)]}{k\cosh [kh]}
+f_{\bf k}\frac{i{\bf k}\cosh [k\eta({\bf r})]}{k^2}
\right]\exp({i{\bf kr}})\,\frac{d^2{\bf k}}{(2\pi)^2}\, d^2{\bf r}
\nonumber\\
&=&\frac{1}{2}\int\left[\psi\hat K\phi
+\eta\nabla\psi\cdot\nabla\phi+\frac{\eta^2}{2}
\nabla\psi\cdot\nabla\hat K\phi
-\frac{\eta^3}{6}\nabla\psi\cdot\nabla(\nabla^2\phi) 
+\dots \right]d^2{\bf r},
\end{eqnarray}
with subsequent substitution
\begin{equation}\label{phi_psi}
\phi\approx\psi-\eta\hat K\psi
+\eta\hat K\eta\hat K\psi+\frac{\eta^2}{2}\nabla^2\psi
-\eta\hat K\left(\eta\hat K\eta\hat K\psi+\frac{\eta^2}{2}\nabla^2\psi
\right)
-\frac{\eta^2}{2}\nabla^2\eta\hat K\psi
+\frac{\eta^3}{6}\nabla^2\hat K\psi.
\end{equation}
The approximate equality (\ref{phi_psi}) follows from an expansion of 
Eq.(\ref{velocity_potential}):
$
\psi\approx\left[1+\eta\hat K-\frac{\eta^2}{2}\nabla^2
-\frac{\eta^3}{6}\nabla^2\hat K\right]\phi.
$
Thus, 
\begin{eqnarray}
{\cal K}&=&\frac{1}{2}\int\Bigg\{\psi\hat K\left[
\psi-\eta\hat K\psi
+\eta\hat K\eta\hat K\psi+\frac{\eta^2}{2}\nabla^2\psi
-\eta\hat K\left(\eta\hat K\eta\hat K\psi+\frac{\eta^2}{2}\nabla^2\psi
\right)
-\frac{\eta^2}{2}\nabla^2\eta\hat K\psi
+\frac{\eta^3}{6}\nabla^2\hat K\psi
\right]\nonumber\\
&&+\eta\nabla\psi\cdot\nabla\left[
\psi-\eta\hat K\psi
+\eta\hat K\eta\hat K\psi+\frac{\eta^2}{2}\nabla^2\psi
\right]+\frac{\eta^2}{2}
\nabla\psi\cdot\nabla\hat K(\psi-\eta\hat K\psi)
-\frac{\eta^3}{6}\nabla\psi\cdot\nabla(\nabla^2\psi) 
+\dots \Bigg\}d^2{\bf r}.
\end{eqnarray}
After simplifying, we obtain 
${\cal H}=\frac{1}{2}\int\left\{\psi\hat K\psi +g\eta^2
+\eta \left[(\nabla \psi)^2-(\hat K\psi)^2\right]\right\} d^2{\bf r}
+{\cal H}^{(4)}+{\cal H}^{(5)} +\dots$,
where
\begin{eqnarray}
{\cal H}^{(4)}&=&\frac{1}{2}\int\left[
\psi\hat K\eta \hat K\eta\hat K\psi 
+\eta^2(\hat K\psi)\nabla^2\psi \right] d^2{\bf r},\label{H_4}
\\
{\cal H}^{(5)}&=&\frac{1}{2}\int\left[ 
\frac{\eta^3}{6}(\hat K\psi)\nabla^2\hat K\psi
- \psi \hat K \eta \hat K \eta \hat K \eta \hat K\psi
-\frac{\eta^3}{3}(\nabla^2\psi)^2
-\eta^2(\hat K \eta \hat K\psi)\nabla^2\psi
-\frac{\eta^2}{2}(\hat K\psi)\nabla^2(\eta \hat K\psi)
\right] d^2{\bf r}.\label{H_5}
\end{eqnarray}
In the same manner, it is also possible to derive the Hamiltonian 
with a higher-order accuracy. 
\end{widetext}

\end{document}